\documentclass[reprint, amsmath,amssymb,aps,floatix]{revtex4-2}

\usepackage{hyperref}
\usepackage{graphicx}
\usepackage{bm}
\usepackage[T1]{fontenc}
\usepackage[utf8]{inputenc}
\usepackage{amsmath}
\usepackage{mathtools}
\usepackage{xcolor}
\usepackage{xspace}
\usepackage{braket}
\usepackage{subfigure}

\begin{document}

\preprint{APS/123-QED}


\title{Solving Nuclear-Structure Problems with the Adaptive Variational Quantum Algorithm
}

\author{A. M. Romero} \email{a.marquez.romero@fqa.ub.edu}
\affiliation{Department of Physics and Astronomy, University of North Carolina,
 Chapel Hill, North Carolina 27516-3255, USA \\ Departament de Física Quàntica i Astrofísica (FQA), Universitat de Barcelona (UB),  c. Martí i Franqués, 1, 08028 Barcelona, Spain \\ Institut de Ciències del Cosmos (ICCUB), Universitat de Barcelona (UB), c. Martí i Franqués, 1, 08028 Barcelona, Spain}

\author{J. Engel} \email{engelj@physics.unc.edu}
\affiliation{Department of Physics and Astronomy, University of North Carolina,
 Chapel Hill, North Carolina 27516-3255, USA}

\author{Ho Lun Tang} \email{holuntang@vt.edu}
\affiliation{Department of Physics, Virginia Tech, Blacksburg, VA 24061}

\author{Sophia E. Economou} \email{economou@vt.edu}  
\affiliation{Department of Physics, Virginia Tech, Blacksburg, VA 24061} 

\date{\today}

\date{\today} 

\begin{abstract}

We use the Lipkin-Meshkov-Glick (LMG) model and the valence-space nuclear shell model to
examine the likely performance of variational quantum eigensolvers in
nuclear-structure theory.  The LMG model exhibits both a phase transition and
spontaneous symmetry breaking at the mean-field level in one of the phases,
features that characterize collective dynamics in medium-mass and heavy nuclei.
We show that with appropriate modifications, the ADAPT-VQE algorithm, a particularly flexible
and accurate variational approach, is not troubled by these complications. We treat 
up to 12 particles and show that the number of quantum operations needed
to approach the ground-state energy scales linearly with the number of
qubits.  We find similar scaling when the algorithm is applied to the nuclear
shell model with realistic interactions in the $sd$ and $pf$ shells. Although most of these simulations contain no noise, we use a noise model from real IBM hardware to show that for the LMG model with four particles, weak noise has no effect on the efficiency of the algorithm.

\end{abstract}

\maketitle

\section{Introduction}

Quantum computers promise to allow quasi-exact solutions of quantum many-body
problems in chemistry and physics without the exponential scaling that plagues
classical methods \cite{mcardle2020}.  Among the many ways of exploiting quantum
computers, hybrid algorithms known as Variational Quantum Eigensolvers
(VQEs)~\cite{peruzzo2014variational, Cerezo2021,tilly2021variational,bharti2021noisy}
which are based on the variational principle of quantum mechanics, are under particularly intensive development. These algorithms allocate optimization of wave functions to
classical computers, using their quantum counterparts only to realize the parameterized states
that the optimization scheme calls for.  The result is
fewer quantum operations (albeit at the expense of more measurements), leading to the hope that near-term quantum circuits,
which are noisy and will be for some time to come, can implement the procedures
without becoming too inaccurate. VQEs have been both tested on existing quantum
processors and simulated classically for a number of simple problems in molecular chemistry~\cite{Kandala2017,mcardle2020,colless2018computation,cao2019quantum}. Despite the potential impact of these algorithms in nuclear structure, much less work has been done in this domain. 

Computations in quantum chemistry and nuclear structure have many similarities, but also important differences. A common approach in the two fields is the application of configuration-interaction methods in which
model spaces are constructed from orbitals that can be empty or occupied.  In
nuclear physics such methods are generically referred to as ``the shell model,''
and range from the diagonalization of phenomenological nucleon-nucleon
interactions in quite limited valence spaces to \textit{ab initio} calculations
with bare nucleon-nucleon interactions in many-shell model spaces with no inert
``core.''  The use of orbitals in nuclear physics, however, may obscure the fact that 
nucleons do not orbit around any fixed points.  The nucleus is self bound, and the 
nucleons that compose it can move in concert without drastically changing their 
total energy.  This low-energy collective motion has many consequences,
the most important of which in the context of VQEs is that mean-field theory, which underlies 
configuration-interaction methods, must spontaneously break symmetries of the Hamiltonian --- 
translational symmetry at least, and sometimes also rotational symmetry, parity, and 
particle-number conservation --- to capture the collective correlations 
corresponding  to shape deformation, superfluidity, etc.
Certain symmetries are broken in some nuclei and
not others, so that a quantum phase transition can occur at critical values of
the neutron and/or proton number.

To assess the performance of VQEs for nuclear physics problems, we must work with relatively simple models and/or systems for which nearly exact solutions
are easy to obtain. 
One such model, which is due to Lipkin, Meshkov, and Glick (the LMG model)~\cite{lipkin1965validity} and
which we will describe 
in detail in the next section, has several virtues.   For certain values of its parameters,
it is a simplified version of a closed-shell nucleus with an isoscalar monopole
giant resonance as an excitation.  When the energy of the resonance goes to zero, the
model exhibits a transition to a ``deformed phase'' \cite{agassi1966validity}
very much like that associated with actual physical deformation.  The symmetry
that is broken in the model is ``number parity,'' which resembles the spatial
parity broken in pear-shaped nuclei.  Finally, the model can be
interpreted as involving interacting spins, which has made it useful for
condensed-matter physics~\cite{unanyan2005many, campbell2016criticality, russomanno2017floquet} and for a benchmark study for quantum chemistry methods~\cite{wahlen2017merging}.  
The second simple case we examine is the shell model itself, under the restriction that both the number of orbitals and the number of nucleons occupying those orbitals are reasonably small.

Two recent papers ~\cite{cervia2021lipkin,Chikaoka_2022} have examined VQEs within the LMG model. 
The authors of Ref.~\cite{cervia2021lipkin} focused on a small number of qubits (up to three) and used an
ansatz 
that enforced the symmetries of the model, so as to search only the relevant 
subspace. Most of the analysis involved running this small version of the problem on quantum hardware and assessing the performance of the hardware when combined with error mitigation techniques. Although the 
symmetry-enforcing circuit was efficent in terms of gate count and number of parameters, it was 
limited to the (exactly solvable) LMG model. Finding an efficient state preparation circuit for larger 
system sizes 
is a nontrivial task, and Ref.~\cite{cervia2021lipkin} cited a CNOT scaling of $\mathcal{O}(2^N)$. 
Ref.~\cite{Chikaoka_2022} compared the unitary coupled cluster ansatz and structure-learning 
ansatz~\cite{Ostaszewski2021structure} through classical simulations of the LMG problem for up to four qubits. 
The performance of this ansatz declined as the interaction strength in the Hamiltonian was increased. 
It therefore remains an open problem to find a suitable VQE approach for the LMG model that scales 
favorably and that can also be generalized to realistic nuclear-structure problems (beyond solvable models) with complex physics such as phase transitions.

Studies of the quantum algorithms in the shell model are fewer.  We are aware only of
Ref.\ \cite{stetcu2021variational}, which analyzed the efficiency of encodings and the 
performance of a unitary coupled clusters ansatz in four nuclei, with up to six 
valence nucleons.

In this paper, we address the challenge of treating the LMG model efficiently in all its complexity, and also  handling the phenomenological shell model, by employing an algorithm known as Adaptive Derivative-Assembled Problem-Tailored VQE (ADAPT-VQE)~\cite{grimsley2019adaptive,tang2021qubit}. The ADAPT-VQE algorithm grows the ansatz iteratively and according to the Hamiltonian that is being simulated. As a result, an ansatz tailored to the problem is created through information obtained by measurements on the quantum computer. We apply ADAPT-VQE to the LMG model on both
sides of the phase transition and then to the phenomenological nuclear shell model with
realistic effective interactions and relatively small numbers of particles. We investigate the scaling of circuit depth with particle number, particularly around the LMG phase transition and when the mean-field 
spontaneously breaks a symmetry. This is crucial to assessing the likely effectiveness of near-term quantum computers.  
We find extremely promising scaling in both cases, even around the phase transition, when we apply 
symmetry-projection techniques from nuclear-structure theory. The problem-tailored nature of the 
ADAPT-VQE ansatz allows the algorithm to adjust the circuit structure and depth according to the demands presented by the problem. 
It also enables easy implementation of additional subroutines for various 
purposes, such as symmetry projection. This makes ADAPT-VQE particularly well suited to problems 
involving a quantum phase transition, including those in nuclear physics. Our results are not 
only important for the quantum simulations of nuclear structure, but also serve as the first test of the ADAPT-VQE algorithm in problems that exhibit complex phenomena such as phase transitions and symmetry breaking.


The article is organized as follows. Section~\ref{sec:lmg} introduces the LMG
model and the mean-field and projection techniques that we use to construct suitable ADAPT-VQE ansatz\"e.
Section~\ref{sec:sm} summarizes the theory of the shell model, which we also implement in ADAPT-VQE. Section~\ref{sec:adapt} briefly presents the
variational quantum algorithm and describes the modifications that we use to include
projection.  Section~\ref{sec:results} presents results on the performance and
scaling of the algorithms, and Sec.~{\ref{sec:conclusions}} offers some
conclusions.


\section{Lipkin-Meshkov-Glick model}\label{sec:lmg}

\begin{figure}
\includegraphics[scale=0.9]{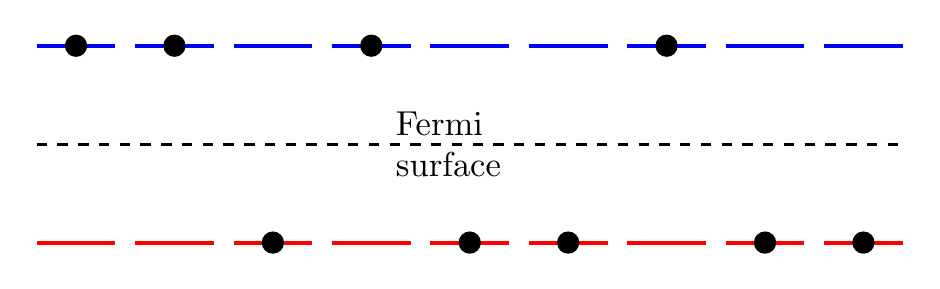}%
\caption{One configuration of the system described by the LMG model with $N=9$.
Particles can only move directly up or down, and each pair of levels (upper and
lower) must contain a total of one particle.  }
\label{fig:lmgconfig}
\end{figure}

The LMG model~\cite{lipkin1965validity} describes a system of $N$
particles moving in two $N-$fold degenerate shells, separated from one another
by a single-particle energy gap as illustrated in Fig~\ref{fig:lmgconfig}.  The
gap mimics a similar gap between nuclear shells, so that the lowest
configuration in the model represents a closed-shell nucleus.  

The LMG Hamiltonian is 
\begin{equation}
\label{eq:H}
    H = t J_z - V (J_x^2 - J_y^2) = t J_z - \frac{V}{2} (J_+^2 + J_-^2) \,,
\end{equation}
where $J_z$ and $J_{\pm} = J_x \pm i J_y$ are generators of an $SU(2)$ algebra
obeying the commutation relations $[J_+,J_-]=2J_z$ and $[J_z,J_{\pm}]=\pm
J_{\pm}$, and are defined in terms of creation and annihilation operators for particles
in the $i^{\rm th}$ lower ($-$) and upper ($+$) levels by the relations 
\begin{equation}
\label{eq:spinops}
    \begin{split}
        J_z &= \frac{1}{2}\sum_i \left( a_{i,+
        }^{\dag} a_{i,+} - a_{i,-
        }^{\dag} a_{i,-}\right) \equiv \frac{1}{2} \sum_{i} \sigma^i_z \,,\\
        J_+ & = \sum_i a_{i,+
        }^{\dag} a_{i,-} \equiv \sum_{i} \sigma^i_+ \,,\\
        J_- & = \sum_i a_{i,-
        }^{\dag} a_{i,+} \equiv \sum_{i} \sigma^i_- \,.
    \end{split}
\end{equation}
The operator $J_{+}$ ($J_-$) raises (lowers) a nucleon from the lower (upper) shell to its counterpart in the upper (lower)
shell, and the operator $J_z$ is the difference between the number of nucleons in
the upper and lower shell.  The form of the coupling in $H$ implies that the
$i^{\rm th}$ lower and $i^{\rm th}$ upper levels must together contain a total
of one nucleon, as illustrated in Fig.\ \ref{fig:lmgconfig}.  If we take $i$ to correspond to single-particle angular-momentum quantum numbers, it also implies that the two-nucleon part of the Hamiltonian
schematically represents the piece that doesn't change a
nucleon's angular momentum; it is this piece that determines the properties of monopole (breathing) resonances.  Finally, as we emphasize by defining the $\sigma$
operators in Eq.\ \eqref{eq:spinops}, it implies that the entire model can be
taken to simulate $N$ interacting spins, with a nucleon in the $i^{\rm th}$
lower (upper) level corresponding to a spinor in its down (up) state.  Spins can
in turn be mapped in a straightforward way to qubits.

Because the eigenstates of $H$ are unchanged by the simultaneous scaling of $t$
and $V$ in Eq.\ \eqref{eq:H}, they really depend only a single parameter.
Fixing the value of the quantity $t+(N-1)V/2$, at 1, we can, without loss of generality in
the eigenvectors, write the LMG Hamiltonian in the form
\begin{equation}
\label{eq:hampauli}
H = \frac{1-y}{2}\sum_i \sigma_z^i - \frac{2y}{N-1}\sum_{i<j}
(\sigma_+^i\sigma_+^j+\sigma_-^i\sigma_-^j) \,, 
\end{equation}
where $y \equiv (N-1)V/2$.  Varying $y$ from 0 to 1 allows us to sample all
possible values for the ratio of the two- and one-body terms in $H$.     

The standard way to get a reasonable approximation to the ground state is
through mean-field theory --- the Hartree-Fock (HF) approximation in our
nucleon-based interpretation.  In the LMG model the HF state always has the
form~\cite{agassi1966validity, ring2004nuclear} 
\begin{equation}\label{eq:slater}
\begin{split}
    \ket{\rm{HF}} &= \prod_i \beta_{i}^{\dag} \ket{\rm{vac}}\\
    & \equiv  \prod_i [\cos(\alpha) a_{i,-}^{\dag} + \sin(\alpha)
    a_{i,+}^{\dag}] \ket{\rm{vac}} \,, 
\end{split}
\end{equation}
where $\ket{\rm vac}$ is the ``bare vacuum,'' the state in which all the levels are unoccupied, and the $\beta^\dag_i$ create quasiparticles that
are superpositions of particles in the lower and upper shells.  The value of
$\alpha$ is obtained by minimizing the expectation value of the energy in the
Hartree-Fock state.  One finds that $\alpha$ vanishes as long as the strength of the interaction, compared to the size of the single-particle splitting, is below a critical value, but becomes nonzero when the strength is above that value, viz.,
\begin{equation}\label{eq:alphadef}
    \cos(2 \alpha) =
    \begin{cases}
    1 & y < \frac{1}{3}  \\
     \frac{1-y}{2y} & y \geq \frac{1}{3} \,.
    \end{cases}
\end{equation}
Thus, a phase transition to non-trivial quasiparticles occurs at $y=1/3$ for
any number of particles $N$.

Instead of transforming the bare vacuum $\ket{\rm vac}$ to the HF state $\ket{\rm HF}$ 
in Eq.\ \eqref{eq:slater}, one can retain the bare vacuum and rotate the operators in
Eq.\ \eqref{eq:hampauli} around the $y$ axis:
\begin{equation}
\label{eq:rotpauli}
\begin{split}
\sigma_x & \longrightarrow \cos(2\alpha) \sigma_x + \sin(2\alpha) \sigma_z, \\  
\sigma_z & \longrightarrow \cos(2\alpha) \sigma_x - \sin(2\alpha) \sigma_z.
\end{split}
\end{equation}
The equivalence makes it easier to manipulate many-quasiparticle states.

Because the LMG Hamiltonian moves particles between the upper an lower shells
only in pairs, it conserves ``number parity,'' the number of particles modulo 2
in the lower shell.  When $y > 1/3$, however, the quasiparticle operators are
superpositions of creation operators for states in both shells, and the state
$\ket{\rm HF}$ breaks number parity spontaneously~\cite{agassi1966validity}.  The
number-parity operator can be written in the form
\begin{equation}
\label{eq:Pi}
    \Pi \equiv  (-1)^{N_{+}} = e^{i\pi (N/2 + J_z)} = (-1)^{N/2}e^{i\pi J_z} \,,
\end{equation}
and has the eigenvalue $+1$ ($-1$) when the system state has an even (odd) number of
particles $N_+$ in the upper shell.  Number-parity symmetry may be restored by
projecting out of $\ket{\rm HF}$ the piece with one or the other value for $\Pi$.
The operator $P_{\pm}$ that projects onto the space of states with even ($+$) or
odd ($-$) parity is
\begin{equation}
\label{eq:projop}
    P_{\pm} = \frac{1}{2} (1\pm \Pi) \,. 
\end{equation}
Like all projectors, these are Hermitian and obey $P_{\pm}^2 = P_{\pm}$.

The restoration of symmetry through projection almost always improves the
accuracy of mean-field approximations in nuclear physics.  Within the LMG
model, the improvement can be explored both analytically
\cite{agassi1966validity} and, when testing approximations such as unitary
coupled clusters that go beyond mean-field theory,
numerically~\cite{harsha2018difference, wahlen2017merging}.  Symmetry breaking
and restoration has also been investigated in applications to quantum
computing~\cite{gard2020efficient, lacroix2020symmetry, guzman2021accessing}.

\section{Nuclear shell model}\label{sec:sm}

\begin{figure}
\includegraphics[scale=1.0]{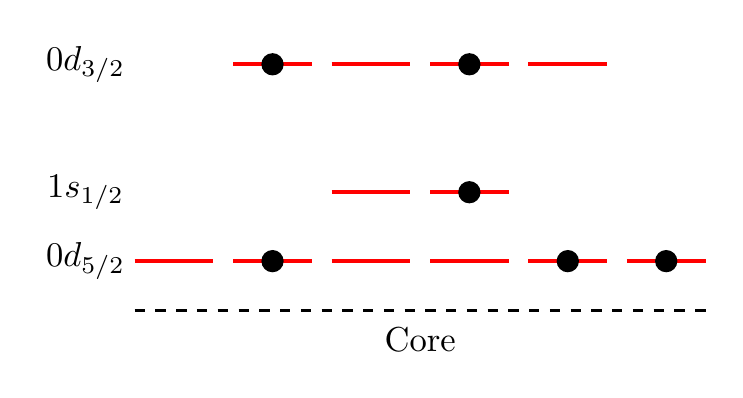}%
\caption{One configuration of a system described by the $sd-$shell valence space, with $N=6$. The orbits are degenerate and the order shown corresponds to that produced by the single-particle part of the USDB Hamiltonian.}
\label{fig:sdfig}
\end{figure}

The shell model is a mainstay of nuclear-structure theory~\cite{caurier2005shell, heyde1994nuclear, de2013nuclear}. It freezes most of the nucleons in an inactive ``core,'' treating only those around the Fremi surface explicitly.  The various many-nucleon product states in this valence space make up a basis in which one represents a nuclear Hamiltonian.  The ground and excited states of the nucleus, along with their energies, are obtained as by direct diagonalization of the Hamiltonian matrix in this effective valence space. The shell model, with a Hamiltonian derived from an underlying nucleon-nucleon interaction and
then tweaked to fit the energies of particular states, is able to accurately
reproduce low-lying spectra (and other properties) of many nuclei. Much like orbital-based models in chemistry, the shell
model has a generic one-plus-two-body Hamiltonian, of the form 
\begin{equation}\label{eq:smham}
    H = \sum_i \epsilon_i a_i^{\dag} a_i + \frac{1}{4} \sum_{ijkl}\bar{v}_{ijkl}
    a_i^{\dag} a_j^{\dag} a_l a_k \,,
\end{equation}
where now $a^{\dag}_i$ $(a_i)$ creates (annihilates) a fermion in orbital $i$, $\epsilon_i$ is the single-particle energy of orbital $i$ and the $\bar{v}_{ijkl} = v_{ijkl} - v_{ijlk}$ are antisymmetrized two-body matrix elements of the internucleon potential. For nuclei
with mass number $A$ between 16 and 40, one typically takes as the valence space
the $sd$-shell, comprising the 0$d_{5/2}$, 0$d_{3/2}$, and 1$s_{1/2}$ orbits, which
amount to a total of 12 single-particle states for both protons and neutrons (a schematic of this valence space containing 6 nucleons is shown in Fig.~\ref{fig:sdfig}).
For somewhat heavier isotopes one often works in the $pf$-shell, comprising the
0$f_{7/2}$, 0$f_{5/2}$, 1$p_{3/2}$, and 1$p_{1/2}$ orbits and amounting to a
total of 20 single-particle states for both protons and neutrons.  
The USDB~\cite{usdb} interaction is the standard two-body Hamiltonian in the $sd$ shell,
and the KB3G~\cite{kb3g} interaction is often used in the $pf$ shell. Ref.\ \cite{stetcu2021variational} used the $sd$ shell in a test of the unitary coupled cluster ansatz that focused on gate depth. 

\section{The ADAPT-VQE algorithm}\label{sec:adapt}

ADAPT-VQE uses an operator pool out of which the trial state is built and a
gradient criterion to determine which operator is appended at every step. 
The operators are successively applied to a reference state (typically the HF
state) and all the parameters are optimized at each step, starting from the
previously optimized values as the initial guess. The authors of
Ref.~\cite{grimsley2019adaptive} simulated the use of the method on a quantum
computer to calculate bond-dissociation curves for the molecules LiH, BeH$_2$,
and H$_6$.  The results were more accurate and required a lower circuit depth
compared to those produced by other variational ans\"atze built from the same
set of operators, such as the widely used unitary coupled cluster singles and
doubles (UCCSD) \cite{lee2018generalized}.  ADAPT-VQE achieves shallower
circuits at the cost of more measurements; considering the noisy nature of
existing and near-term hardware, this trade-off is advantageous.

To analyze the performance of ADAPT-VQE in both the LMG model and the nuclear
shell model, we simulate its operation on a classical computer.  The quantum
algorithm starts with a reference state $ \ket{\rm ref}$ and approaches the
ground-state through the successive application of unitary operators,
constructed by exponentiating simple excitation operators $A_{i}$ from a
predefined pool,
\begin{equation}
\label{eq:genstate}
    \ket{n} =  e^{i \theta_{n} A_{n}} \ket{n-1} = \prod_{k=1}^n e^{i \theta_k
    A_k} \ket{\rm ref} \,.
\end{equation}
Here the $\theta$'s are parameters with values that produce the minimum possible
energy, and our convention for the product, as in Ref.\
\cite{grimsley2019adaptive}, is $\prod_{k=1}^n \mathcal{O}_k \equiv
\mathcal{O}_n \dots \mathcal{O}_1$ so that order of operators is reversed from
that in the usual convention.  All parameters are optimized after the
application of each operator, so that $\ket{n}$ is not necessarily related
simply to $\ket{n-1}$.  The optimization procedure is based on the fact that the
derivative of the energy at iteration $n$ with respect to a parameter in the
ansatz is the expectation value of the commutator of the corresponding pool
operator with the Hamiltonian~\cite{grimsley2019adaptive},
\begin{equation}
\label{eq:gradients}
\frac{\partial E^{(n)}}{\partial \theta_k} = i \braket{ n | [H, A_k] | n} \,. 
\end{equation}
The algorithm selects as $A_{n + 1}$ the operator that produces the largest
derivative in Eq.\ \eqref{eq:gradients}; to do so it relies on the quantum
circuit to construct the states in Eq.\ \eqref{eq:genstate}, from which the
derivatives can be constructed by measuring the commutators.  The new values for
the $n+1$ parameters $\theta_k$ are then obtained by minimizing of the total
energy on a classical computer with measurements of the energy on the quantum
computer guiding the minimization.

To keep the circuit depth small, we restrict ourselves to a pool consisting of
one- and two-body operators $A_i$, each of which acts on particles in at most
two pairs of levels.  In the LMG model, the one-body operators acting on the
particles in level pair $k$ are
\begin{equation}
\label{eq:onebodyops}
\begin{split}
    S_+^k & = \sigma_+^k + \sigma_-^k=\sigma_x^k, \\ 
    S_-^k & = -i(\sigma_+^k - \sigma_-^k)=\sigma_y^k ,\\ 
    S_0^k & = \sigma_z^k \,, 
\end{split}
\end{equation}
where we have used the ladder Pauli operators
$\sigma_{\pm}=\frac{1}{2}(\sigma_x\pm i \sigma_y)$. For the two-body operators,
with $j<k$, the pool is
\begin{equation}\label{eq:twobodyops}
\begin{split}
T_+^{jk} & = \sigma_+^j \sigma_+^k + \sigma_-^j \sigma_-^k = \frac{1}{2}(\sigma_x^j\sigma_x^k-\sigma_y^j\sigma_y^k), \\ 
T_-^{jk} & = -i(\sigma_+^j \sigma_+^k - \sigma_-^j \sigma_-^k) = \frac{1}{2}(\sigma_x^j\sigma_y^k+\sigma_y^j\sigma_x^k),\\ 
U_+^{jk} & = \sigma_+^j \sigma_-^k + \sigma_-^j \sigma_+^k = \frac{1}{2}(\sigma_x^j\sigma_x^k+\sigma_y^j\sigma_y^k),\\
U_-^{jk} & = -i(\sigma_+^j \sigma_-^k - \sigma_-^j \sigma_+^k) = \frac{1}{2}(\sigma_y^j\sigma_x^k-\sigma_x^j\sigma_y^k),\\
V_+^{jk} & = (\sigma_+^j + \sigma_-^j) \sigma_z^k=\sigma_x^j\sigma_z^k,\\
V_-^{jk} & = -i(\sigma_+^j - \sigma_-^j)\sigma_z^k = \sigma_y^j\sigma_z^k,\\
V_0^{jk} & = \sigma_z^j\sigma_z^k.\\
\end{split}
\end{equation}
The pool thus contains $3N$ one-body operators and $7N(N-1)/2$ two-body
operators of each kind, exhausting all the possible Hermitian combinations of
Pauli operators.  

The algorithm's reference state $\ket{\rm ref}$, the wave function at iteration
zero, can be chosen freely.  We use two different initial states for the LMG
model: the single-configuration uncorrelated state $\ket{0}$, in which all
particles are in the lower shell (or all the spins are down in the spin-model
interpretation) and the mean-field Hartree-Fock state state $\ket{\rm HF}$,
which spontaneously breaks number-parity symmetry for $y>1/3$.  These two states
are the same for $y<1/3$ and differ for $y>1/3$.  We can also modify the pool
operators by building them from the HF quasiparticle operators $\beta^\dag_i$
and $\beta_i$ in Eq.\ \eqref{eq:slater} rather than directly from the particle
and hole operators.  The use of the HF initial state and the HF quasiparticle
operator pool together, is equivalent to using $\ket{\rm 0}$ and the ordinary
particle-hole operator pool, with a Hamiltonian transformed according to
\eqref{eq:rotpauli}.

In addition to choosing a LMG reference state, we can choose whether or not to
use symmetry projection at various points in the algorithm, most easily through
the ``projected Hamiltonian'' $P_+ H P_+$ (see Appendix \ref{sec:php}).  To
differentiate the combinations of reference states and kinds of projection that
we employ in the LMG model, we use the following naming scheme for what we call
``methods:'' 
\begin{itemize}
   \item 0 - The reference state is the uncorrelated one, $\ket{\rm 0}$, and the
   operator pool contains the usual one- and two-particle-hole excitation
   operators.
   \item HF - $\ket{\rm HF}$ is used as the initial state, together with the
   quasiparticle operator pool and no symmetry projection.
   \item HF-PAV - Same as HF except that the projected Hamiltonian $P_+HP_+$ is
   monitored to assess convergence rather than $H$ itself. The acronym PAV,
   which stands for ``projection after variation,'' comes from nuclear-structure
   theory.
   \item HF-VAP - Same as HF-PAV, but the projected Hamiltonian is also used to
   evaluate the gradients in Eq.\ \eqref{eq:gradients} and to minimize the
   cost-function.  This procedure is close to ``variation after projection'' in
   nuclear-structure theory. 
\end{itemize}

ADAPT-VQE is perfectly able to handle the shell-model
Hamiltonians~(\ref{eq:smham}) as well. Because the Hamiltonian is more general
than the LMG interaction, the representation of the system in terms of qubits is
more involved.   Here, we will use the Jordan-Wigner
mapping~\cite{jordan1993paulische, ortiz2001quantum} between the fermonic and
Pauli operators
\begin{equation}
    \begin{split}
        a_i^{\dag} &= \Bigg(\prod_{k=0}^{i-1}\sigma_z^k\Bigg)\sigma_-^i, \\
        a_i &= \Bigg(\prod_{k=0}^{i-1} \sigma_z^k\Bigg)\sigma_+^i.
    \end{split}
\end{equation}
The operator pool contains all possible two-body fermion operators $a_i^{\dag}
a_j^{\dag} a_l a_k$, where $i<j$ and $l<k$ are single-particle labels. Thus,
they are of the form 
\begin{equation}
    T_{rs}^{pq} = i (a_p^{\dag} a_q^{\dag} a_r a_s - a_r^{\dag} a_s^{\dag} a_p a_q),
\end{equation}
where antisymmetrization has been taken into account explicitly.  Although
shell-model Hamiltonians can break symmetries, the conserved quantities are more
complicated than number-parity and we will not examine the effects of
shell-model projection here. 

\section{Results}\label{sec:results}

\subsection{LMG model}

\begin{figure}
\includegraphics[width=0.45\textwidth]{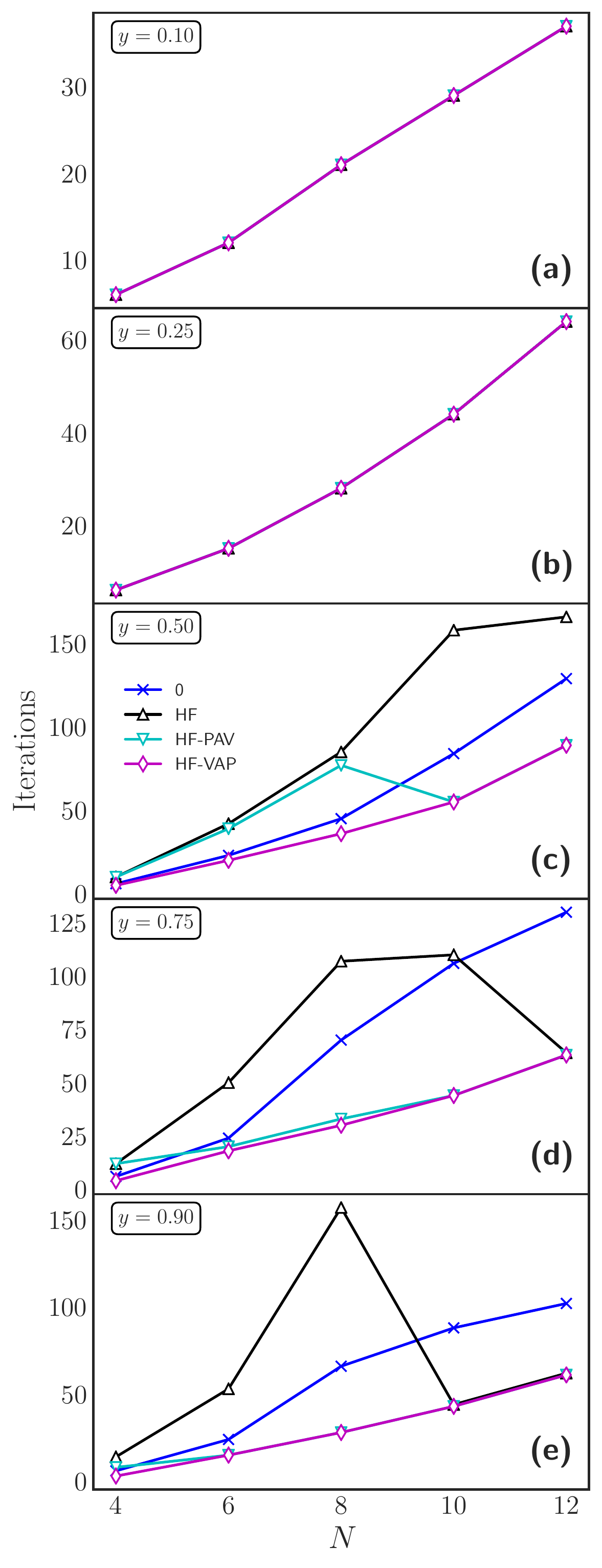}%
\caption{Number of iterations required for ADAPT-VQE to come within $0.1\%$ of the LMG
ground-state energy, as a function of the number of particles $N$ and for
several values of the coupling constant $y$ in the
Hamiltonian~(\ref{eq:hampauli}).  Each iteration corresponds to the presence of
an additional pool operator in the wave function.}
\label{fig:scaling}
\end{figure}

For quantum computers to be useful in the near term in nuclear physics, we need
algorithms in which circuit depth increases mildly with particle number and/or
model-space size.  With our algorithm, this depth is related to the number of
pool operators needed to approximate the exact ground state well.  In Fig.\
\ref{fig:scaling}, we plot the number of such operators (or, equivalently, the
number of parameters) required to obtain the ground-state energy to within
$0.1\%$ as a function of the number of qubits $N$ for several values of $y$ and
all the algorithm variants outlined in the previous section.  For $y<1/3$, in
the symmetry-unbroken phase, the methods are all equivalent.  For $y>1/3$, the
performance of different methods diverges.  First, we can see that until the
number of particles is sufficiently large, method HF is not as good even as
method 0, particularly around the phase transition point $y=1/3$ where the
mean-field approach fails.  The two symmetry-projecting methods perform the
best, except for small numbers of particles, where method 0 slighly outperforms
method HF-PAV.  Method HF-VAP always gives the best performance.  The difference
between HF-VAP and HF-PAV shows that using the projected Hamiltonian for
constructing the ansatz improves the algorithm significantly.  Projection in the
computation of gradients is easy for ADAPT-VQE because of its flexibility; the
cost-function used in the choosing operators can be modified according to the
features of the problem being solved.

Once $y$ and $N$ are large enough, the symmetry-projecting methods do not
perform substantially better than the method HF.  In addition, method 0, which
never breaks symmetry, does considerably worse than those that do at large $y$.
For such ``strongly deformed'' systems, the crucial thing is to include
important correlations in the reference state by breaking number-parity
symmetry. Restoring the symmetry afterwards is less helpful; mean-field theory
is all that is needed in this regime.  Adding symmetry projection reduces the
difficulties that mean-field theory encounters around the phase transition
without affecting its performance for large $y$ and large system size.  The
combination provides a universal scheme, useful for the whole range of $y$ and
$N$.

\begin{figure}
\includegraphics[width=0.45\textwidth]{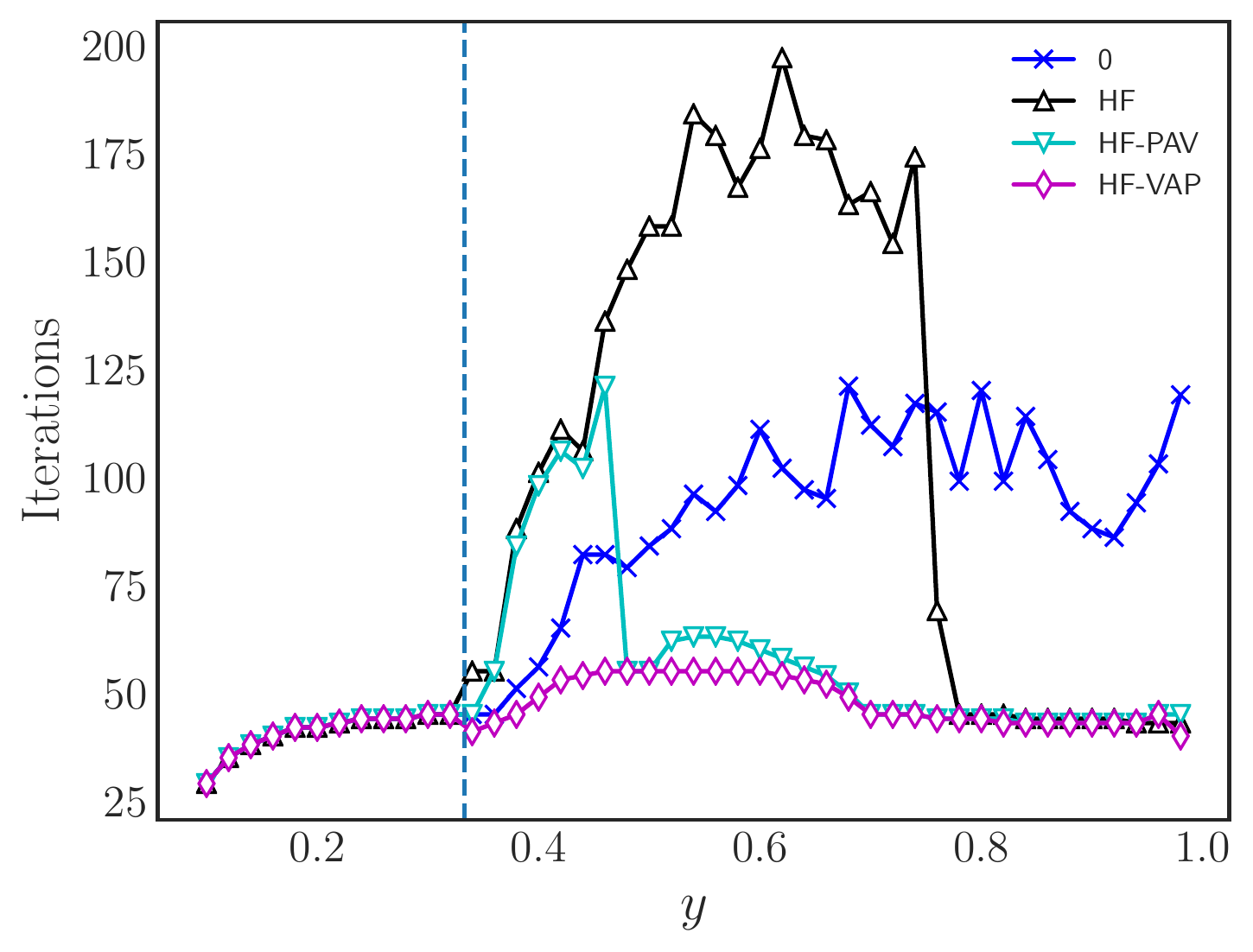}%
\caption{Same as Fig.\ \ref{fig:scaling} but for a fixed number of particles
($N=10$) and as a function of the coupling $y$.  }
\label{fig:yplot}
\end{figure}

\begin{figure}
\includegraphics[width=0.45\textwidth]{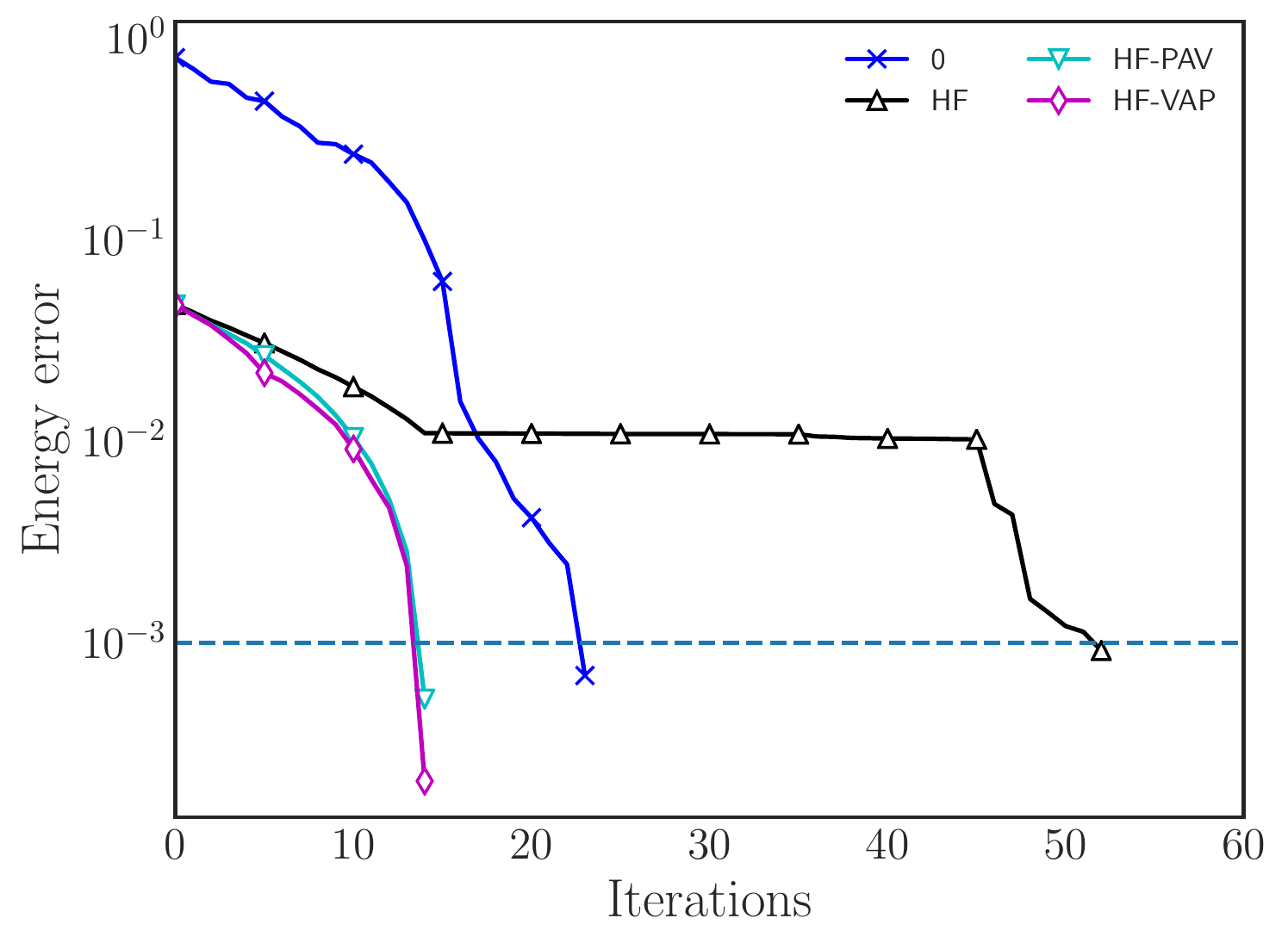}%
\caption{Error in the ratio of the ADAPT energy to the exact one for the LMG model as a function
of the number of pool operators in the ADAPT ground state, for $N=6$ and $y =
0.9$.  Method HF features a long plateau before the quick drop.} 
\label{fig:enerconv}
\end{figure}

In Fig.\ \ref{fig:yplot}, one can see irregularities in the curves produced by
methods 0 and HF.  These have to do with the criteria for convergence of the
algorithm.  The addition of a single operator to a chain that produces the
system's approximate state can have almost no effect or, occasionally, a large
effect.  If the large effect happens to reduce the energy enough so that it
satisfies the convergence criterion, the iteration ends.  If the effect is not
quite large enough, the iteration continues and may not end until much later,
when another significant reduction occurs.   Fig.~\ref{fig:enerconv} illustrates
this phenomenon.  If iterations were halted when the error in the ratio of the
energy to the exact one reached a little over $10^{-2}$ instead of $10^{-3}$,
the HF methods would require only about 15 operators instead of 50.  In reality,
of course, we don't know the exact energy and so have to truncate when the
energy appears stable. As the figure shows, and as Ref.\
\cite{grimsley2019adaptive} notes, long plateaus can then cause the algorithm to
terminate too soon. Modifying the convergence criterion could alleviate this
problem.

Figure \ref{fig:scaling} (d) and (e) show, in addition to the overall scaling
already discussed, a non-monotonic trend for the HF energy as a function of the
number of particles. The peak at $N=8$ is due to a convergence plateau; at
larger $N$, such a plateau is encountered only later, closer to the exact
ground-state energy. We believe that the plateau stems from an inefficient pool
or initial state (because of symmetry violation in this instance) and that the
algorithm therefore needs more parameters to escape from the plateau (see the
results for method HF in Fig.\ \ref{fig:enerconv}).  The reason for the
lower-energy plateau at larger $N$ could be the increased efficiency of
mean-field theory there.

\begin{figure}
\includegraphics[width=0.45\textwidth]{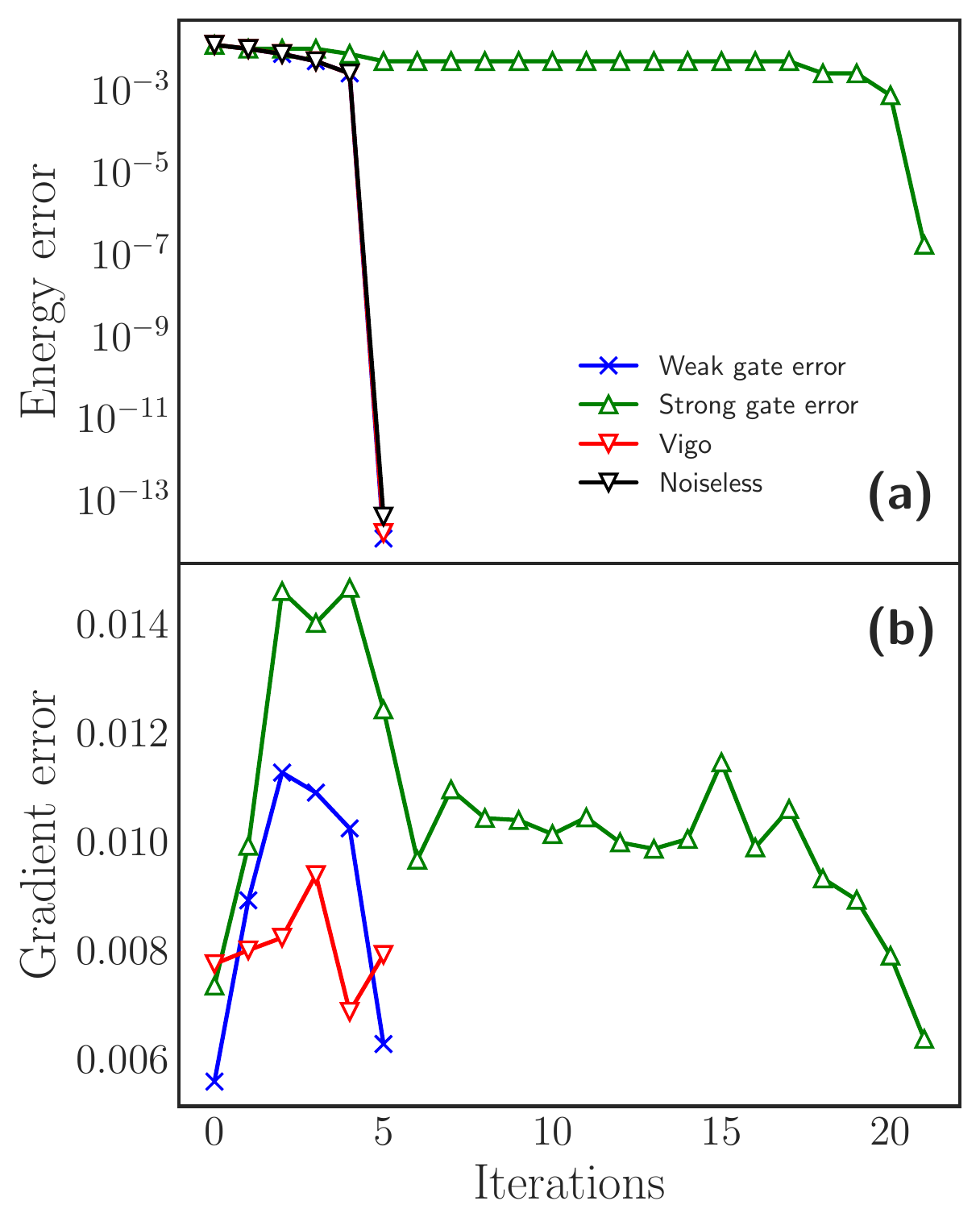}%
 \caption{Effects of noise in a simulation of ADAPT for the LMG model with $N=4$
 and $y=0.3$, and with the uncorrelated reference state $\ket{0}$.  All
 measurements were simulated with 8000 shots. In (a), the energy error is
 plotted as a function of the number of operators in the ansatz. The ``weak gate
 error'' label correspond to the custom noise model with depolarizing error
 rates for the single and two-qubit gates of 10\% and 20\%; those rates for the
 ``strong gate error'' case are 20\% and 40\%. The Vigo, weak noise, and
 noiseless results are identical.  In (b), the averaged deviation of the
 measured derivatives from the noiseless values throughout the algorithm is
 plotted as a function of the number of operators in the ansatz. The errors
 increase with the number of iterations but only to a point, after which they
 decrease.  The reason for the eventual decrease is that the gradient itself
 also eventually decrease. The ``strong gate error'' results show a faster
 increase in gradient error as the circuit grows because the gate error
 accumulates faster. }
        \label{fig:noise}
\end{figure}

One might expect ADAPT to be vulnerable to noise because it relies on
measurements of energy gradients, which in practice are affected by
imperfections in the device and controls.  To address the issue of robustness
against noise, we ran noisy simulations with the built-in noise model in Qiskit
\cite{Qiskit}, using both real noise data from the IBM quantum device Vigo, a
quantum processor that consists of five transmons connected in a T-shaped
layout, and a custom noise model.  The simulation with real noise data contains
gate depolarizing error, measurement error and shot noise due to finite sample
size.  Only gate depolarization and shot noise are included in our custom model,
and we vary the gate-error rate in the model.  To focus on gradient measurement,
which is the distinguishing feature of ADAPT-VQE, we simulated the standard VQE
part of the calculation noiselessly.  The rate of convergence in the energy for
the simulation with real noise and our custom model with weak noise (see Fig.
\ref{fig:noise} and its caption for details) are the same as in the noiseless
simulation.  This comparison shows that the algorithm is accurate as long as the
noise level is below a certain threshold, i.e.\ that ADAPT ansatz-construction
algorithm is robust. Though the effects of noise should eventually be explored
in more detail, these results are promising for ADAPT-VQE.

\subsection{Nuclear shell model}

In this subsection, we apply ADAPT-VQE to valence-space shell-model
Hamiltonians, for isotopes in the $sd$ and $pf$ shells.  We use the USDB and
KB3G interactions mentioned in Sec.~\ref{sec:sm} here (with no mass-dependent
modifications) together with the configuration-interaction code
BIGSTICK~\cite{bsI,bsII} whose exact results serve as a benchmark.  In the $sd$
shell, we consider isotopes of oxygen, which has no valence protons, and neon,
which has two. In the $pf$ shell we examine isotopes of calcium, which has no
valence protons.  Nuclei more complicated than those are a large burden for
simulations.  We increase the neutron number until the shell is half full;
further increases reduce the dimension of the Hilbert space.

Good mean-field theory in the shell model would involve pair correlations and
particle-number-violating (not number-parity-violating) reference states, and so
we limit ourselves to an analog of method 0, choosing the configuration --- a
set of filled and empty orbitals --- with the lowest average energy and choosing
randomly when we encounter degeneracy.  We thus violate no symmetries and need
no projection.  A thorough investigation of mean-field symmetry breaking in this
context will be the focus of future work.

\begin{figure}
\includegraphics[width=0.45\textwidth]{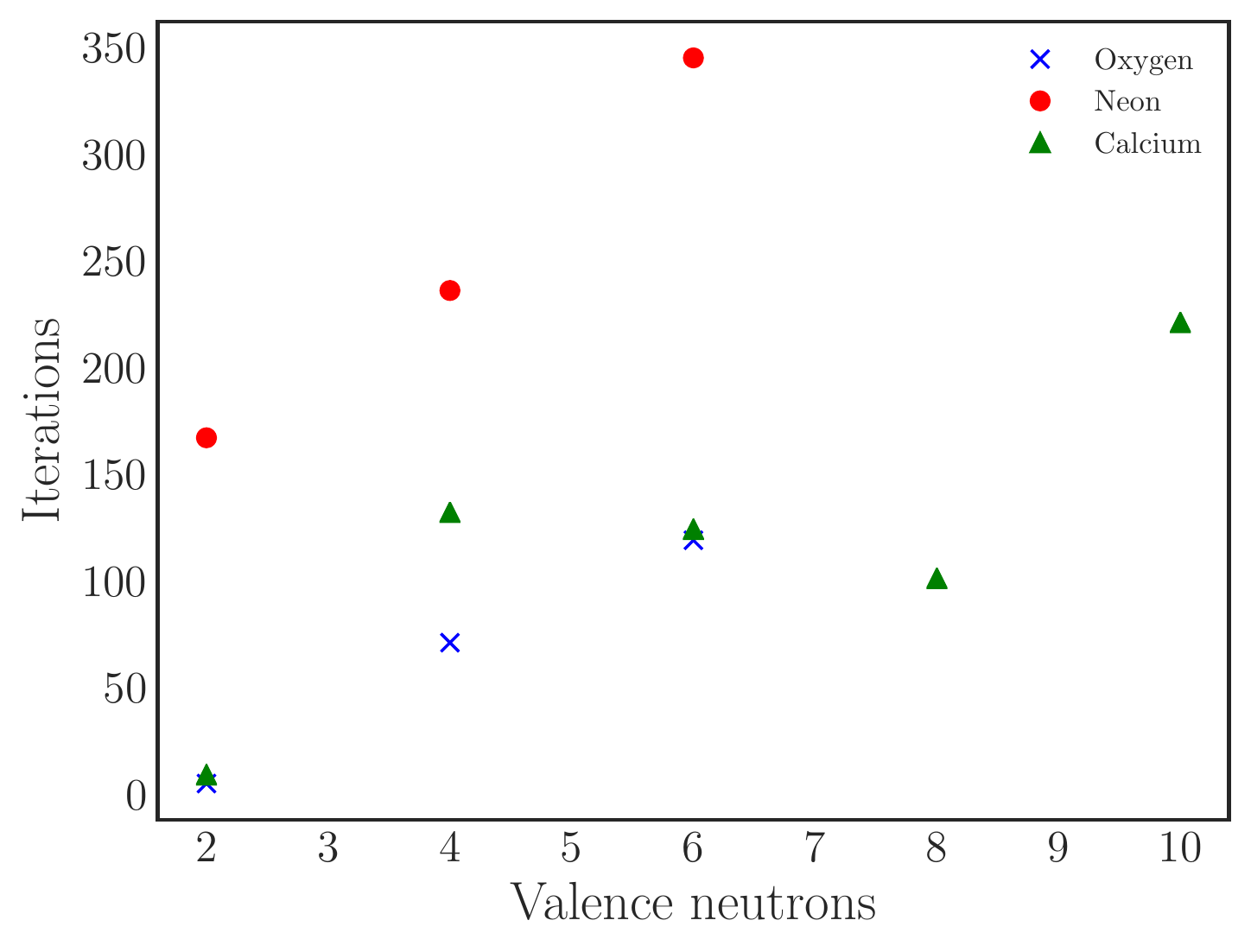}%
\caption{Number of operators needed to reach the ground-state energy of the nucleus
to within 1\%, 2\% and $0.0001$\% for isotopes of calcium, neon and oxygen,
respectively, as a function of the number of valence neutrons in the shell.}
\label{fig:shellmodelfig}
\end{figure}

Figure \ref{fig:shellmodelfig} shows how the number of operators needed to
reproduce the ground-state energies in these isotopes scales with particle
number. Though the protons in neon make it more complicated than oxygen, the
trend is clearly linear in both $sd$-shell isotopic chains.  In calcium, a
$pf$-shell chain, the sequence of points is not monotonic, for the same reasons
as in the LMG model (see Fig.~\ref{fig:scaling}), but the rate of increase
overall is low.  Although we have not analyzed noise in this context, the mild
scaling both here and in the LMG model is extremely promising for ADAPT-VQE.  

\section{Conclusions}\label{sec:conclusions}  
In this work, we have investigated the question of whether nuclear-structure physics, which encompasses collective motion,
phase transitions, and complicated correlations, can potentially benefit from near-term quantum
computers. By examining the scaling of the performance of ADAPT-VQE, a problem-tailored
variational quantum algorithm that dynamically creates the ansatz, in both a model-problem with many features of real nuclei
and in realistic valence-space shell-model calculations, we conclude that the
benefits could be substantial.  We find mild scaling in all cases, including
near a phase transition, if we first allow our reference state to break
symmetries and then restore those symmetries before measuring energies.

The scaling we 
have demonstrated here, 
both away from the phase transition and near it, combined with a level of noise robustness in the construction of the variational ansatz, is 
encouraging for the future of VQEs in nuclear physics.

\section{Acknowledgments}

This work was supported in part by the U.S. Department of Energy Office of
Science. J.E.\ acknowledges support from the DOE Office of Nuclear Physics,
under grant No.\ DE-FG02-97ER41019. S.E.E.\ acknowledges support from the DOE
National Quantum Information Science Research Centers, Co-design Center for
Quantum Advantage (C2QA), contract number DE-SC0012704.  Computing resources
were provided by the Research Computing group at the University of North
Carolina and the National Energy Research Scientific Computing Center (NERSC), a
U.S.\ Department of Energy Office of Science User Facility located at Lawrence
Berkeley National Laboratory, operated under Contract No.\ DE-AC02-05CH11231.

\appendix

\section{Computation of Projected Hamiltonian and Gradient}
\label{sec:php}

We write the LMG Hamiltonian in Eq.\ \eqref{eq:hampauli} in the form 
\begin{equation}
H = \frac{1-y}{2}\sum_i \sigma_z^i - \frac{y}{N-1}\sum_{i<j}
(\sigma_x^i\sigma_x^j-\sigma_y^i\sigma_y^j)  \,. 
\end{equation} 
The parity projection operator is defined by Eqs.\ \eqref{eq:Pi} and
\eqref{eq:projop}.   Using
\begin{equation}
e^{i\pi J_z} = e^{i\frac{\pi}{2} \sum_j \sigma_z^j} = i^N \prod_j
     \sigma_z^j \,, 
\vspace{.02in}
\end{equation}
we obtain the projected Hamiltonian $P_+ H P_+ \equiv H P_+$,
\begin{equation}
\begin{aligned}
H P_+ & = \frac{1-y}{4} \sum_{i}  \left( \sigma_z^i + \prod_{j \neq
    i} \sigma_z^j\right) - \frac{1}{2}\frac{y}{N-1} \\
& \times \sum_{i<j}
\left(\sigma_x^i \sigma_x^j 
 - \sigma_y^i\sigma_y^j + \prod_{k\neq i,j}
    \sigma_x^k \sigma_z^i \sigma_z^j- \prod_{k\neq i,j} \sigma_y^k \sigma_z^i
    \sigma_z^j\right) \,.
\end{aligned}
\end{equation}
We rotate the $\sigma$ operators in this expression as in Eq.\
\eqref{eq:rotpauli} to apply the VAP and PAV methods. 

Method VAP requires minimizing the projected energy,
\begin{equation}
  E_{\phi} = \frac{\braket{\phi| H P_+|\phi}}{\braket{\phi |P_+|\phi}} \,,
\end{equation}
where $\ket{\phi}$ is the ansatz for the state $\ket{n}$ in Eq.\
\eqref{eq:genstate}.  The supplementary information for Ref.\
\cite{grimsley2019adaptive} shows that without the projectors one has 
\begin{equation}
    \frac{\partial E_{\phi}}{\partial \theta_i} = 2 \operatorname{Re} \braket{
    \sigma_i^H |T_i|\psi_{1,i}}\,, 
\end{equation}
where 
\begin{equation}
\label{eq:vardefs}
\begin{aligned}
T_k &= i A_k \,, \\
\ket{\sigma_i^{\mathcal{O}}} &= \prod_{j=N}^{i+1} \exp(-\theta_k T_j)
\mathcal{O} \ket{\phi} \,, \\
\ket{\psi_{1,i}} &= \prod_{j=1}^{i} \exp(\theta_j T_j) \ket{0} \,, 
\end{aligned}
\end{equation}
and, as always, the usual product convention is reversed.  In our case, with
projectors included, this expression becomes
\begin{equation}
\begin{split}
\frac{\partial E_{\phi}}{\partial \theta_i} = \frac{1}{|\braket{ \phi
    |P_+|\phi}|^2} \Big( & 2 \operatorname{Re} \braket{ \sigma_i^{HP_+}
    |T_i|\psi_{1,i}} \braket{ \phi |P_+|\phi} - \\ & 2 \operatorname{Re}
    \braket{ \sigma_i^{P_+} |T_i|\psi_{1,i}} \braket{ \phi |HP_+|\phi} \Big) \,. 
\end{split}
\end{equation}


%

\end{document}